\documentclass[pre,twocolumn]{revtex4}
\usepackage{epsfig}
\usepackage{amsmath}
\usepackage{hyperref}
\usepackage{breakurl}

\begin{document}

\title{Helical distributed chaos in rotating turbulence and convection \\ 
(with applications to geomagnetic dynamo)}

\author{A. Bershadskii}

\affiliation{
ICAR, P.O. Box 31155, Jerusalem 91000, Israel}

\begin{abstract}

 The role of the moments of helicity distribution in rotating turbulence has been studied using the notion of helical distributed chaos. Results of the direct numerical simulations, laboratory experiments and geophysical observations have been used in this investigation. It is shown, in particular, that even for the cases when the global helicity is equal to zero at least even moments are usually non-zero (due to the appearance of the local spatial regions with strong negative and positive helicity) and can play a significant role in the rotating turbulence as adiabatic invariants. Rotating buoyancy driven thermal convection (Rayleigh-B\'{e}nard and Rayleigh-Taylor, the former also for the magnetohydrodynamics) is also studied, and applications of this approach to the convection zone of massive stars and to the geomagnetic dynamo have been discussed in this context.
\end{abstract}

\maketitle

\section{Introduction}

\begin{figure} \vspace{-0.5cm}\centering
\epsfig{width=.45\textwidth,file=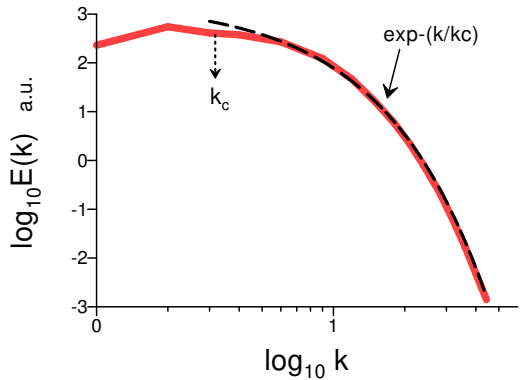} \vspace{-3.8cm}
\caption{Energy spectrum of the freely decaying rotating turbulence at a large viscosity.}
\end{figure}

  Rotating turbulence is very important for engineering, geophysics, and astrophysics. The problems related to the rotating turbulence are numerous and still are far from a comprehensive understanding (see for a recent review Ref. \cite{bif}). Numerical simulations and laboratory experiments indicate that helicity can play a significant role in rotating turbulence. Even in the cases when the net helicity is zero (due to a global symmetry) the local regions with strong negative and positive helicity are usually observed in the rotating flows. However, a thorough and universal theoretical understanding of how it works is still missing. One should begin from the deterministic chaos in the rotating fluids to develop such understanding.

  It is well known that the chaotic smooth dynamical systems with compact strange attractors have exponential temporal (frequency) spectra \cite{oh}-\cite{mm}.  For the dynamical systems described by the equations with partial derivatives (as in fluids dynamics), the spatio-temporal smoothness should also result in the spatial (wavenumber) exponential spectra \cite{mm},\cite{kds}.\\

  The incompressible Navier-Stokes equations 
$$
\frac{\partial {\bf u}}{\partial t} + {\boldsymbol \omega} \times
  {\bf u} + 2 {\boldsymbol \Omega} \times {\bf u}  = 
  - \nabla {\tilde p} + \nu \nabla^2 {\bf u} + {\bf f}  \eqno{(1)}
$$
$$  
  \nabla \cdot \bf{u}=0   \eqno{(2)}
$$  
are usually used in the direct numerical simulations (DNS) of the rotating turbulence. Here ${\bf u}$ and $ {\boldsymbol \omega}$ are the velocity and vorticity fields, ${\tilde p}=p+|{\bf u}|^2/2$ is a modified pressure, $\nu$ is a kinematic viscosity, ${\boldsymbol \Omega}$ is an imposed solid body rotation, ${\bf f}$ is an external force.\\

  In the paper Ref. \cite{bokh} a direct numerical simulation using the Eqs. (1-2) were performed for unbounded freely decaying (${\bf f} =0$) homogeneous turbulence with imposed solid body rotation at constant angular velocity. Namely, the background solid body rotation was suddenly imposed upon a well developed isotropic turbulence in a spatial domain with the periodic boundary conditions. \\ 
  
  Figure 1 shows energy spectrum of the freely decaying rotating turbulence at the time $t_f$ - the end of the anisotropic computation with the viscosity $\nu = 1/600$ (in the terms of the Ref. \cite{bokh}, the spectral data were taken from Fig. 4a of the Ref. \cite{bokh}).  This viscosity is the largest one from those used in the DNS and one can expect that the rotating turbulence was decaying enough to be reduced to a deterministic chaos state. Indeed, the dashed curve is drawn in the Fig. 1 to indicate the exponential wavenumber spectrum

$$
 E(k) \propto \exp-(k/k_c) \eqno{(3)}
$$ 

Position of the characteristic wavenumber $k_c$ is shown in the Fig. 1 by a dotted arrow.

\section{Helicity dynamics}

   For the inviscid case the equation for the mean helicity is
$$
\frac{d\langle h \rangle}{dt}  = 2\langle {\boldsymbol \omega}\cdot (-2 {\boldsymbol \Omega} \times {\bf u} + {\bf f})  \rangle \eqno{(4)} 
$$ 
here $h={\bf u}\cdot {\boldsymbol \omega}$ is the helicity distribution field, and $\langle...\rangle$ is an average over the spatial volume. One can conclude from the Eq. (4) that the mean helicity generally is not an inviscid invariant in this case or it is equal to zero due to a global spatial symmetry.  In this paper, we will consider the cases when the large-scale motions only contribute the main part to the correlation $\langle {\boldsymbol \omega}\cdot (-2 {\boldsymbol \Omega} \times {\bf u} + {\bf f}  \rangle$. The correlation $\langle {\boldsymbol \omega}\cdot (-2 {\boldsymbol \Omega} \times {\bf u} + {\bf f}  \rangle$, however, is swiftly decreasing with spatial scales (this is typical for the chaotic and turbulent flows). Therefore, despite the mean helicity is not generally an inviscid invariant the higher moments of the helicity distribution can be still considered as inviscid invariants \cite{lt},\cite{mt}.\\

   To show this, let us divide the spatial volume into a net of the cells which are moving with the fluid (Lagrangian description) - $V_i$ \cite{lt}\cite{mt}. The boundary conditions on the surfaces of the subdomains are taken as ${\boldsymbol \omega} \cdot {\bf n}=0$. Moments of order $n$ can be then defined as 
 $$
I_n = \lim_{V \rightarrow  \infty} \frac{1}{V} \sum_j H_{j}^n  \eqno{(5)}
$$
here the total helicity $H_j$ in the subvolume $V_j$ is
$$
H_j = \int_{V_j} h({\bf r},t) ~ d{\bf r}.  \eqno{(6)}
$$
   Due to the swift reduction of the correlation $\langle {\boldsymbol \omega}\cdot (-2 {\boldsymbol \Omega} \times {\bf u} + {\bf f}  \rangle$ with the scales the helicities $H_j$ can be still approximately considered as inviscid invariants for the cells with the small enough spatial scales. These cells supply the main contribution to the moments  $I_n$ with $n \gg 1 $ for the strongly chaotic flows (cf. \cite{bt}).  Hence, the moments $I_n$ with the sufficiently large $n$ can be still considered as inviscid quasi-invariants despite the total helicity $I_1$ cannot. For the strongly chaotic (turbulent) flows even the value $n=2$ can be considered as reasonably large  (the moment $I_2$ is the Levich-Tsinober invariant of the inviscid turbulence\cite{lt}). In the inertial range of scales such moments can be considered as {\it adiabatic} invariants for the viscous cases. \\

    The basins of attraction of the chaotic attractors corresponding to the adiabatic invariants $I_n$ can be considerably different. Namely, the chaotic attractors corresponding to the smaller $n$ have thicker basins of attraction (a kind of the intermittency). Therefore, the dynamics of the flow is dominated by the adiabatic invariant $I_n$ with the smallest order $n$. \\
   
   Let us begin our consideration from  $I_3$, for simplicity. With decreasing of the viscosity the parameter $k_c$ in the Eq. (3) becomes fluctuating. The dimensional considerations can be used for estimation of the characteristic velocity $u_c$ for the fluctuating $k_c$ 
 $$
 u_c \propto |I_3|^{1/6} k_c^{1/2}    \eqno{(7)}
 $$
in this case

\section{Helical distributed chaos}

     On the other hand the fluctuations of the $k_c$ can be taken into account using an ensemble average 
$$
E(k) \propto \int_0^{\infty} P(k_c) \exp -(k/k_c)~ dk_c  \eqno{(8)}
$$
where the $P(k_c)$ is the probability distribution of the characteristic scale $k_c$. \\
 
    If we assume a Gaussian distribution of the characteristic velocity $u_c$  \cite{my}, we can obtain the distribution $P(k_c)$ for the helically dominated distributed chaos from the Eq. (7)
$$
P(k_c) \propto k_c^{-1/2} \exp-(k_c/4k_{\beta})  \eqno{(9)}
$$
with a new constant parameter $k_{\beta}$ (cf. Eq. (10)).

    Substituting the Eq. (9) into the Eq. (8) we obtain
$$
E(k) \propto \exp-(k/k_{\beta})^{1/2}  \eqno{(10)}
$$     
 
 for the helical distributed chaos.\\

  It is natural that for the {\it smooth} dynamical systems the power spectra have the stretched exponential form. The stretched exponential Eq. (10) can be generalized
$$
E(k) \propto \exp-(k/k_{\beta})^{\beta},   \eqno{(11)}
$$ 
where the parameter $k_{\beta}$ is a new constant. 
  
  In the general case, to find the value of the parameter $\beta$, one can use asymptotic properties of the distribution $P(k_c)$  (at $k_c \rightarrow \infty$).  For this asymptotic, it follows from the Eqs. (8) and (11) that \cite{jon}
$$
P(k_c) \propto k_c^{-1 + \beta/[2(1-\beta)]}~\exp(-bk_c^{\beta/(1-\beta)}), \eqno{(12)}
$$
here $b$ is a constant. The asymptotic distribution $P(k_c)$ can be also found from the dimensional considerations. Namely, the estimate Eq. (7) can be generalized as
$$
 u_c \propto |I_n|^{1/2n}~ k_c^{\alpha_n}    \eqno{(13)}
 $$    
with
$$
\alpha_n = 1-\frac{3}{2n}  \eqno{(14)}
$$  

 If the characteristic velocity $u_c$ has Gaussian distribution the exponents $\beta_n$ and $\alpha_n$ are related  (from the Eqs. (12) and (13)) by the equation
$$
\beta_n = \frac{2\alpha_n}{1+2\alpha_n}  \eqno{(15)}
$$

\begin{figure} \vspace{-1.6cm}\centering
\epsfig{width=.45\textwidth,file=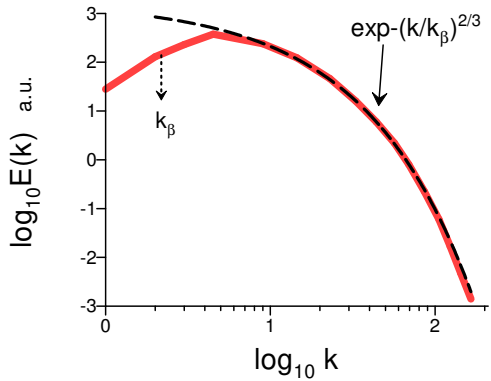} \vspace{-3.6cm}
\caption{As in Fig. 1 but for a smaller viscosity.}
\end{figure}
\begin{figure} \vspace{-0.5cm}\centering
\epsfig{width=.45\textwidth,file=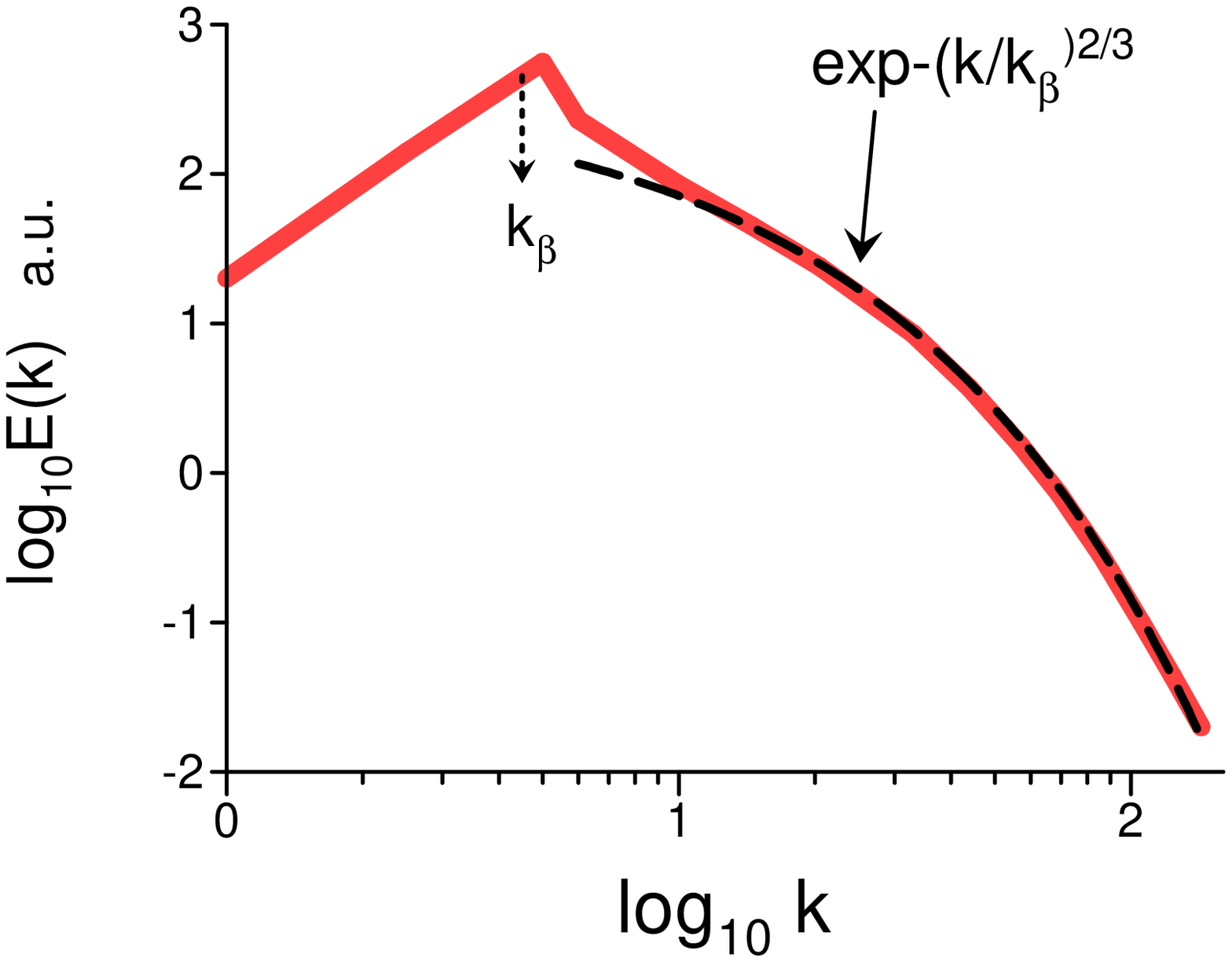} \vspace{-3.9cm}
\caption{Spherically averaged energy spectrum obtained with a non-helical large-scale forcing at $Re_{\lambda} =111$.}
\end{figure}
\begin{figure} \vspace{-1.2cm}\centering
\epsfig{width=.45\textwidth,file=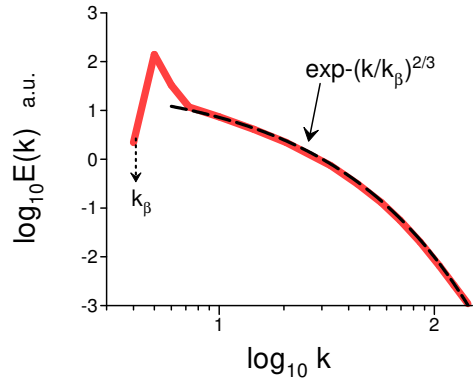} \vspace{-4.05cm}
\caption{The same as in Fig. 3 but obtained with a helical large-scale forcing.  }
\end{figure}

  Then, substituting the value of $\alpha_n $  from the Eq. (14) into the Eq. (15) one obtains
 $$
 \beta_n = \frac{2n-3}{3n-3}   \eqno{(16)}  
 $$
 
     For $n \gg 1$ we obtain from Eq. (16) $\beta_n \simeq 2/3$ i.e., 
$$
E(k) \propto \exp-(k/k_{\beta})^{2/3}  \eqno{(17)}
$$ 
  
  It should be noted that for the case when the net helicity is equal to zero (i.e., $I_1 =0$), at least the even moments $I_n$ with $n \geq 2$ generally are non zero (finite), due to the appearance of the local spatial regions with strong negative and positive helicity \cite{mt}.\\

  For strong chaotization of the helicity fields the value $n=2$ can be treated as sufficiently large to consider the $I_2$ (the Levich-Tsinober invariant of the Euler equation \cite{lt},\cite{mt}) in the inertial range as an adiabatic invariant. In this case we obtain from the Eq. (16) the $\beta =1/3$, i.e.,
$$
E(k) \propto \exp-(k/k_{\beta})^{1/3}  \eqno{(18)}
$$ 

  For $n=4$ the $\beta_n =5/9$ i.e.,
$$
E(k) \propto \exp-(k/k_{\beta})^{5/9}  \eqno{(19)}
$$

\section{Direct numerical simulations} 

  Figure 2 shows energy spectrum of the freely decaying rotating turbulence at a considerably smaller viscosity $\nu = 1/3000$ (in the terms of the Ref. \cite{bokh}, the spectral data were taken from Fig. 4c of the Ref. \cite{bokh}), then that shown in the Fig. 1.  The dashed curve indicates the stretched exponential spectral law Eq. (17). One can see that at this value of viscosity the deterministic chaos (with the exponential spectrum) has been replaced by the distributed chaos (with the stretched exponential spectrum).\\
 
\begin{figure} \vspace{-1.4cm}\centering
\epsfig{width=.47\textwidth,file=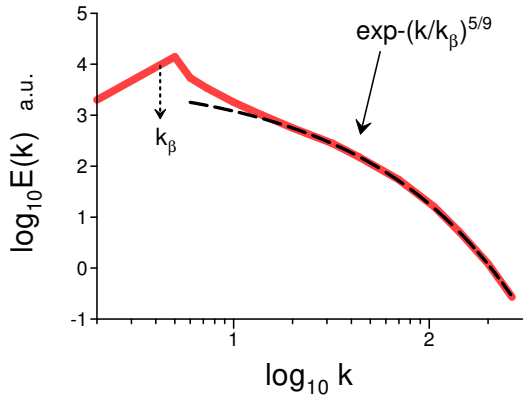} \vspace{-4.1cm}
\caption{Spherically averaged energy spectrum obtained with a non-helical large-scale forcing at $Re_{\lambda} =187$.}
\end{figure}
 
\begin{figure} \vspace{-0.3cm}\centering
\epsfig{width=.45\textwidth,file=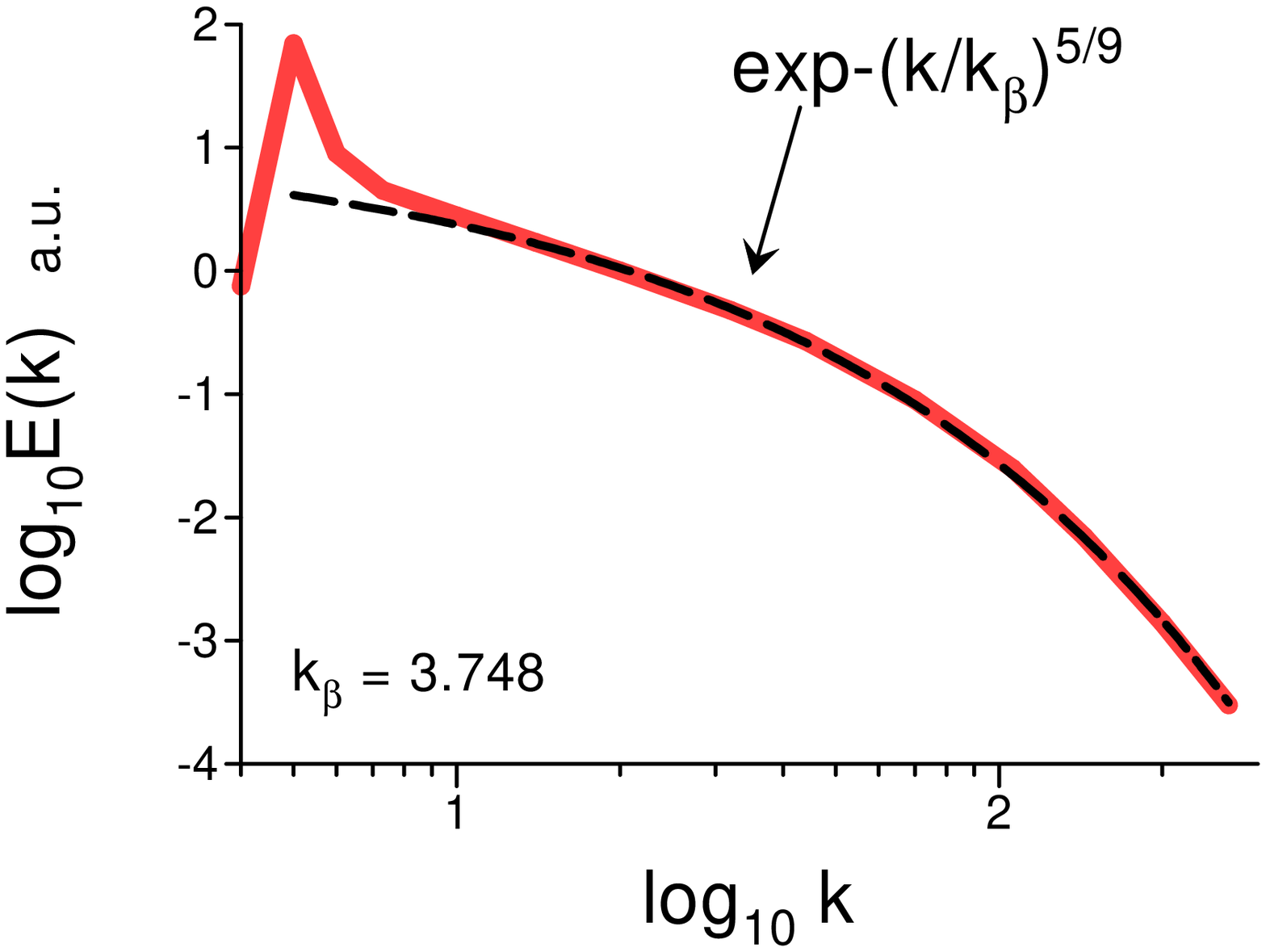} \vspace{-3.6cm}
\caption{The same as in Fig. 5 but obtained with a helical large-scale forcing}
\end{figure}
\begin{figure} \vspace{-0.5cm}\centering
\epsfig{width=.45\textwidth,file=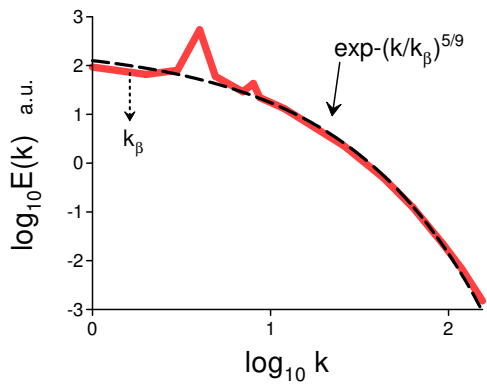} \vspace{-3.6cm}
\caption{ Energy spectrum parallel to the axis of rotation ($k$ is the wavevector parallel to the axis of rotation).}
\end{figure}
  
  In the freely decaying turbulence the Reynolds numbers are rather restricted and to reach higher Reynolds numbers one should use an external forcing ${\bf f}$. Figure 3 shows spherically averaged energy spectrum obtained in recent paper Ref. \cite{vng} for a DNS of the rotating incompressible fluid Eqs. (1-2) with a non-helical large-scale forcing ${\bf f}$ (the Euler scheme) at the Taylor-microscale Reynolds number $Re_{\lambda} =111$ and the Rossby number $Ro =0.206$. The DNS was performed in a $2\pi$ periodic 3D cube. Despite the forcing was non-helical the strong negative and positive helicity was developed in the local regions. The spectral data were taken from Fig. 13 ($R^1_{nh}$) of the Ref. \cite{vng}. The dashed curve indicates the stretched exponential spectral law Eq. (17) (cf. Fig. 2). 
  
  Figure 4 shows analogous spectrum obtained at close values of $Re_{\lambda} =116$ and $Ro =0.228$ but for a helical large-scale forcing ${\bf f}$ (the Euler scheme). The spectral data were taken from Fig. 13 ($R_h$) of the Ref. \cite{vng}. The dashed curve indicates the same stretched exponential spectral law Eq. (17) (cf. Figs. 2 and 3).\\
  
  Let us consider an additional increase in the Reynolds number. Figure 5 shows spherically averaged energy spectrum obtained in the Ref. \cite{vng} with a non-helical large-scale forcing at the Taylor-microscale Reynolds number $Re_{\lambda} =187$ and the Rossby number $Ro =0.216$. The spectral data were taken from Fig. 18 ($S^2_{nh}$) of the Ref. \cite{vng}. The dashed curve indicates the stretched exponential spectral law Eq. (19). 
  
  Figure 6 shows analogous spectrum obtained at close values of $Re_{\lambda} =193$ and $Ro =0.240$ but for a helical large-scale forcing ${\bf f}$ (the Euler scheme). The spectral data were taken from Fig. 18 ($R_h$) of the Ref. \cite{vng}. The dashed curve indicates the same stretched exponential spectral law Eq. (19). \\

   Figure 7 shows energy spectrum parallel to the axis of rotation (where $k$ is the wavevector parallel to the axis of rotation) observed in a direct numerical simulation reported in the paper Ref. \cite{map} at the Reynolds number $Re = 1100$ and the Rossby number $Ro= 0.07$ (a strong rotation). In this simulation performed in a $2\pi$ periodic 3D cube a coherent (the Taylor-Green) non-helical forcing was used at intermediate scales: the   forced wavenumber $k_f \simeq 6.9$. The net helicity in this case is equal to zero but strong negative and positive helicity was observed in the local regions. The spectral data were taken from Fig. 7 (B3) of the Ref. \cite{map}. The dashed curve indicates the stretched exponential spectral law Eq. (19). 
   
   Figure 8 shows corresponding energy spectrum perpendicular to the axis of rotation (where $k$ is the wavevector perpendicular to the axis of rotation). The spectral data were taken from Fig. 5 (B3) of the Ref. \cite{map}. The dashed curve indicates the stretched exponential spectral law Eq. (18). \\
  
  Figure 9 shows energy spectrum perpendicular to the axis of rotation (where $k$ is the wavevector perpendicular to the axis of rotation) obtained in an analogous DNS but for $Re = 5000$ and $Ro = 0.03$, also the forced wavenumber was considerably smaller $k_f = \sqrt{3}$. The spectral data were taken from Fig. 1 of the Ref. \cite{leo}. The dashed curve indicates the stretched exponential spectral law Eq. (18).\\

\section{Laboratory experiments and the Taylor's hypothesis}

  In the laboratory experiments the velocity fluctuations are usually measured with the spatially localized probes and corresponding time series are used to compute 
frequency spectra. The commonly used Taylor's `frozen' turbulence hypothesis assumes that the variability in the measured by a spatially localized probe temporal velocity fluctuations is dominated by the advection of the spatial structures of the velocity field across the local point (the probe) of the measurements rather than their temporal local variability. Accordingly, this hypothesis suggests that the computed from the obtained time series frequency spectrum $E(f)$ should be transformed into a wavenumber spectrum $E(k)$ by transformation $E(k) = UE(f)/2\pi$ with $k = 2\pi f/U$, where $U$ is a mean velocity of the fluid passing the probe (see for a recent review and discussion of the hypothesis applicability Ref. \cite{kv}).\\

\begin{figure} \vspace{-1.5cm}\centering
\epsfig{width=.45\textwidth,file=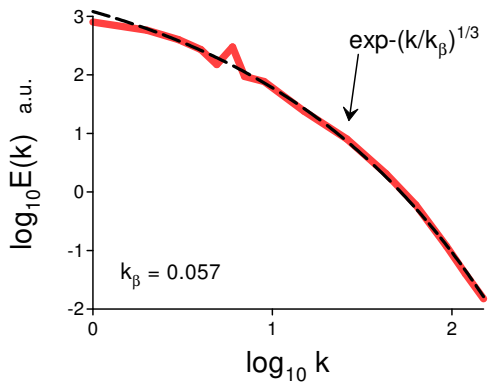} \vspace{-3.9cm}
\caption{Energy spectrum perpendicular to the axis of rotation ($k$ is the wavevector perpendicular to the axis of rotation).}
\end{figure}
\begin{figure} \vspace{-0.1cm}\centering
\epsfig{width=.45\textwidth,file=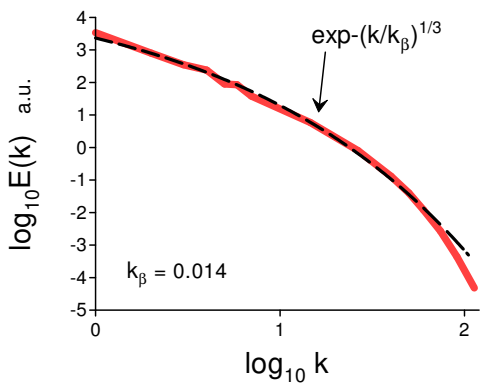} \vspace{-4cm}
\caption{ Energy spectrum perpendicular to the axis of rotation ($k$ is the wavevector perpendicular to the axis of rotation), but for considerably larger $Re$ and smaller $Ro$ (i.e. for a stronger turbulence and rotation than that shown in the Figure 8). }
\end{figure}
\begin{figure} \vspace{-1.7cm}\centering
\epsfig{width=.45\textwidth,file=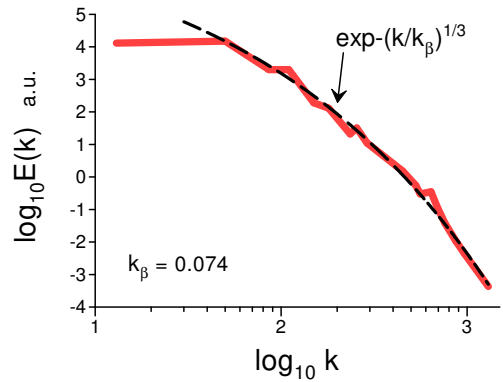} \vspace{-3.7cm}
\caption{ Power spectrum of the time and azimuthal averaged zonal velocity field (here $k$ is the radial wavenumber).} 
\end{figure}
\begin{figure} \vspace{-0.1cm}\centering
\epsfig{width=.45\textwidth,file=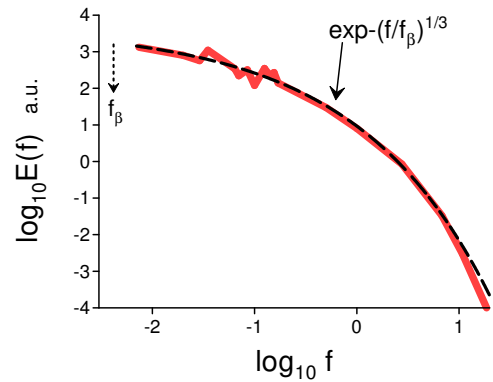} \vspace{-4cm}
\caption{ Frequency power spectrum of the zonal and radial components of the velocity field measured by 7500 probes located in a horizontal plane.}
\end{figure}
 
  To check applicability of the Taylor's hypothesis to rotating turbulence let us use results of the DNS reported in recent paper Ref. \cite{cfb}. In this DNS a rapidly rotating and highly turbulent incompressible fluid was studied in a cylindrical geometry with bottom and side no-slip boundary conditions, and a top flat stress-free surface. The Taylor-Green small-scale ($k_f =12$) forcing was used at this DNS. The net helicity in this case is equal to zero (non-helical forcing) but strong negative and positive helicity was observed in the local regions. The Reynolds number $Re = 4000$ and the Rossby number $Ro =0.002$.\\ 
  
   Figure 10 shows wavenumber power spectrum of the time and azimuthal averaged zonal velocity field ($k$ is the radial wavenumber). The spectral data were taken from Fig. 3a (DNS2) of the Ref. \cite{cfb}. The dashed curve indicates the stretched exponential spectral law Eq. (18).\\
   
   Figure 11 shows frequency power spectrum of the zonal and radial components of the velocity field measured by 7500 probes located in a horizontal plane. The spectral data were taken from Fig. 5b of the Ref. \cite{cfb}. The dashed curve indicates the stretched exponential spectral law Eq. (18) (see the above described Taylor's hypothesis).\\

   Now let us turn to the laboratory experiments. In recent paper Ref. \cite{sal} the rotating turbulence were studied in a cylindrical geometry where two co-rotating (with the same angular velocity) propellers at the bottom and the top of the cylinder produced one large vortex inside the cylinder. The fluid was liquid helium-4 at a `classical' temperature $T= 2.5 K$, i.e. a classical fluid described by the Navier–Stokes Eqs. (1-2). Since the viscosity of this fluid is very small the Reynolds numbers were rather large: $Re = 1.1 \times 10^6$ at $\Omega = 0.19$ rad/s, and $Re = 2.3 \times 10^7$ at $\Omega = 3.77$ rad/s. The hot-wire probes were placed at mid-height in the equatorial plane of the above described co-rotating  Von K\'{a}rmán flow.\\
   
      Figure 12 shows power spectrum of the velocity filed measured at $Re = 1.1 \times 10^6$ ($\Omega = 0.19$ rad/s). The spectral data were taken from Fig. 9 of the Ref. \cite{sal}. Using the Taylor's hypothesis we can compare the spectrum with the Eq. (18) (the dashed curve). From the position of $f_{\beta}$ (the vertical arrow) one can see that the large-scale coherent structures control entire distributed chaos in this case.
      
\begin{figure} \vspace{-1.5cm}\centering
\epsfig{width=.45\textwidth,file=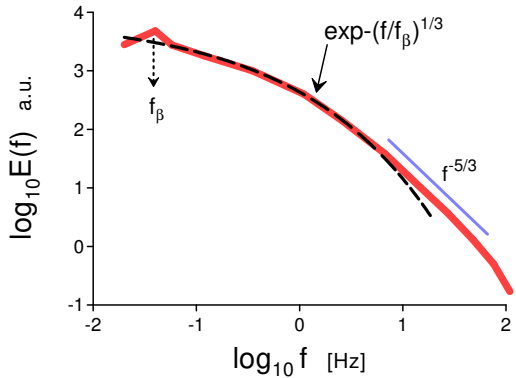} \vspace{-3.6cm}
\caption{Power spectrum of the velocity filed measured at $Re = 1.1 \times 10^6$ ($\Omega = 0.19$ rad/s).}
\end{figure}
\begin{figure} \vspace{-1.3cm}\centering
\epsfig{width=.45\textwidth,file=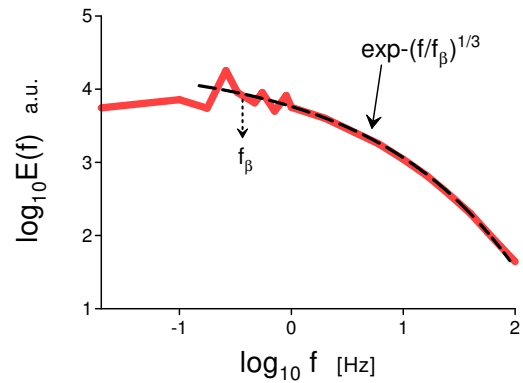} \vspace{-3.8cm}
\caption{As in Fig. 12 but for $Re = 2.3 \times 10^7$ ($\Omega = 3.77$ rad/s).} 
\end{figure}
\begin{figure} \vspace{-0.5cm}\centering
\epsfig{width=.45\textwidth,file=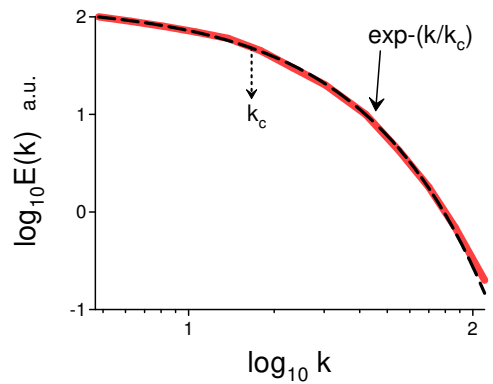} \vspace{-3.95cm}
\caption{Kinetic energy spectrum of the baroclinic convective eddies' background at the end  of the computations.  } 
\end{figure}
    Figure 13 shows power spectrum of the velocity filed measured at $Re = 2.3 \times 10^7$ ($\Omega = 3.77$ rad/s). The spectral data were taken from Fig. 9 of the Ref. \cite{sal}. Using the Taylor's hypothesis we can compare the spectrum with the Eq. (18) (the dashed curve).

\section{Rotating buoyancy driven convection}

   In the most of the geophysical and astrophysical applications the rotating turbulence is considered together with buoyancy driven thermal convection.\\

The simplest buoyancy driven thermal convection under an imposed solid body rotation is described in the Boussinesq approximation by the system of equations (cf. Eqs. (1-3))

 $$
\frac{\partial {\bf u}}{\partial t} {\boldsymbol \omega} \times
  {\bf u} + 2 {\boldsymbol \Omega} \times {\bf u}  =  - \nabla {\tilde p} + \sigma g T {\bf e}_z + \nu \nabla^2 {\bf u}   \eqno{(20)}
$$
$$
\frac{\partial T}{\partial t} + ({\bf u} \cdot \nabla) T  =   \kappa \nabla^2 T, \eqno{(21)}
$$
$$
\nabla \cdot \bf u =  0 \eqno{(22)}
$$
where $T$ is the temperature field, ${\bf e}_z$ is a unit vector along the gravity direction, $g$ is the gravity acceleration,  $\kappa$ is the thermal diffusivity and $\sigma$ is thermal expansion coefficient of the fluid.  \\    

\begin{figure} \vspace{-1.2cm}\centering
\epsfig{width=.45\textwidth,file=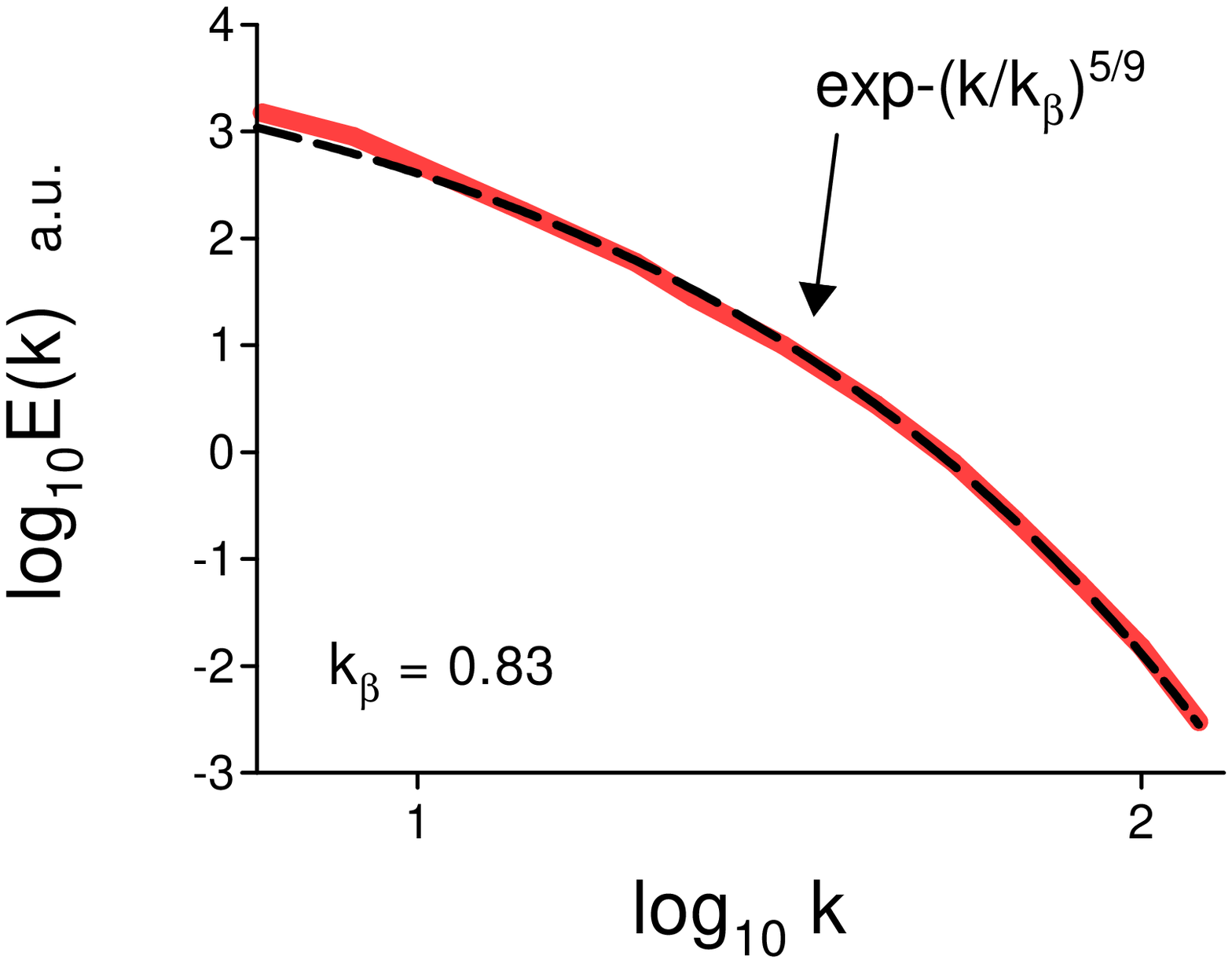} \vspace{-3.9cm}
\caption{Kinetic energy spectrum of the barotropic convective eddies' background at the end  of the computations. } 
\end{figure}

\begin{figure} \vspace{-0.3cm}\centering
\epsfig{width=.45\textwidth,file=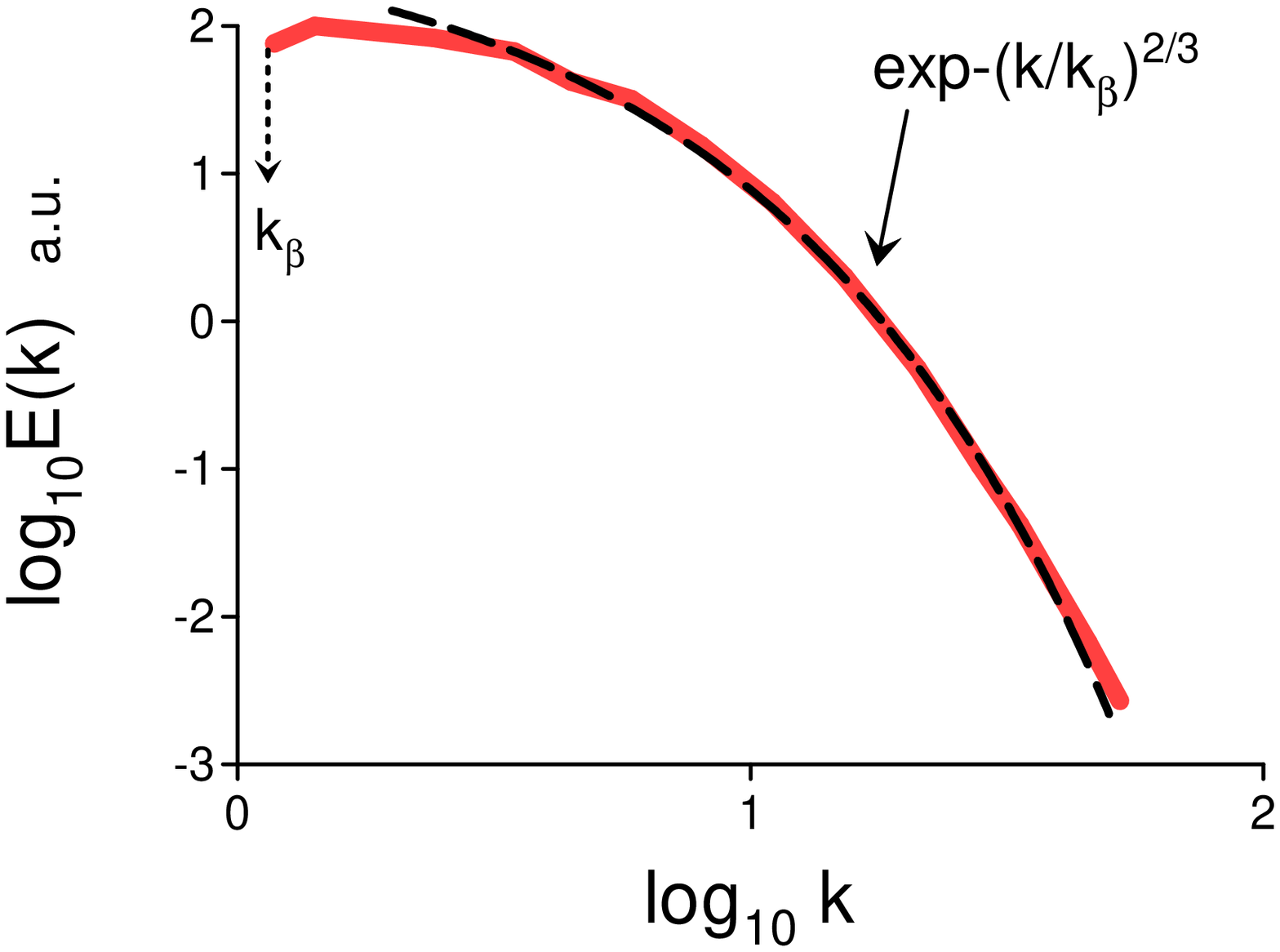} \vspace{-4cm}
\caption{Kinetic energy spectrum obtained for a non-rotating convection zone.} 
\end{figure}
\begin{figure} \vspace{-0.3cm}\centering
\epsfig{width=.45\textwidth,file=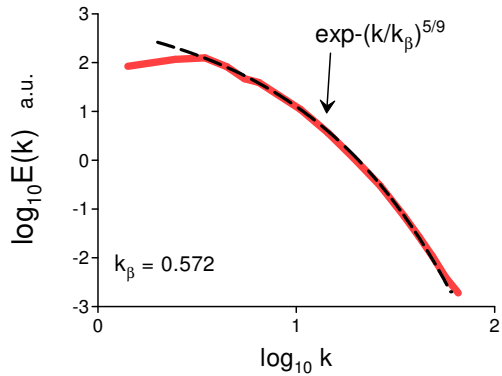} \vspace{-4cm}
\caption{Kinetic energy spectrum obtained for a rotating convection zone. } 
\end{figure}

  If we replace the external force ${\bf f}$ in the Eq. (4) by the thermal convective forcing $\sigma g T {\bf e}_z$, then we obtain
$$
\frac{d\langle h \rangle}{dt}  = 2\langle {\boldsymbol \omega}\cdot (-2 {\boldsymbol \Omega} \times {\bf u} + \sigma g T {\bf e}_z)  \rangle \eqno{(23)} 
$$ 
 and all the considerations of the Sections II and III can be applied to this case as well.    \\

 \subsection{Rayleigh-B\'{e}nard convection}
 
   Let us start from the rapidly rotating turbulent thermal (Rayleigh-B\'{e}nard) convection. A remarkable property of such convection is formation of the large-scale barotropic (depth-independent) vortices, the so-called spectral condensation, even in the 3D case (see, for instance, Ref.  \cite{rub} and references therein). These large-scale vortices take their energy from a background consisting of  the small-scale convective eddies and, on the other hand, provide a corresponding organization of the small-scale convective eddies' background (a positive feedback loop). \\ 
   
   Figure 14 shows kinetic energy spectrum of the baroclinic convective eddies' background at the end of the direct numerical simulations reported in the Ref. \cite{rub} ($t=100$ in the terms of the Ref. \cite{rub}). The spectral data were taken from Fig. 3 of the Ref. \cite{rub}. The DNS were performed in a heated from below spatial domain with periodic in the horizontal directions boundary conditions, impenetrable boundary conditions in the vertical directions, and in the rapid rotation limit $Ro\rightarrow 0$. The dashed curve indicates the exponential spectral law Eq. (3) (i.e. {\it deterministic} chaos). The value $k_c = 20$ (the dotted arrow in the Fig. 14) precisely corresponds to the convective energy injection wavenumber in the terms of the Ref. \cite{rub}, i.e. the convective energy ejection controls the baroclinic deterministic chaos in this case.\\ 
\begin{figure} \vspace{-1.6cm}\centering
\epsfig{width=.45\textwidth,file=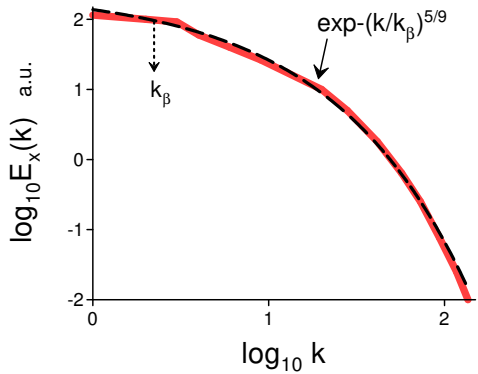} \vspace{-3.5cm}
\caption{Power spectrum of the horizontal component of the velocity - $E_x$, against the wavenumber $k = \sqrt{k_x^2+k_y^2+k_z^2}$ at $\Omega =0$.} 
\end{figure}
\begin{figure} \vspace{-0.5cm}\centering
\epsfig{width=.45\textwidth,file=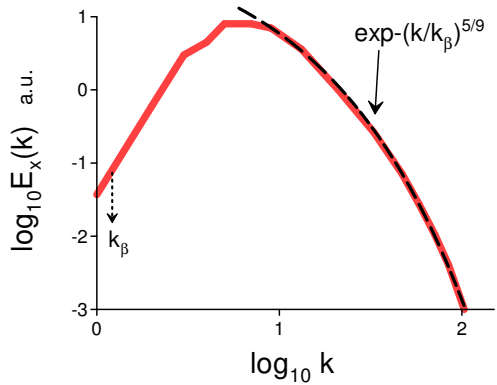} \vspace{-3.2cm}
\caption{Power spectrum of the horizontal component of the velocity - $E_x$, at $\Omega \tau =12.5$.} 
\end{figure}
\begin{figure} \vspace{-0.5cm}\centering
\epsfig{width=.45\textwidth,file=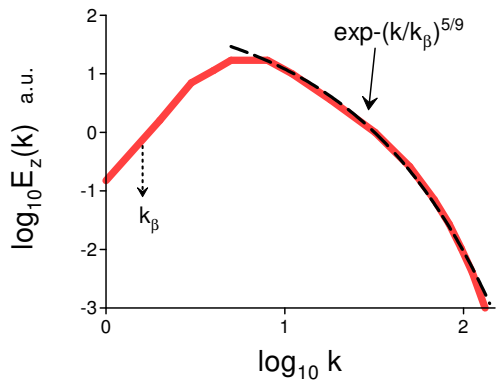} \vspace{-3.8cm}
\caption{As in Fig. 19 but for the vertical component of the velocity field. } 
\end{figure}
\begin{figure} \vspace{-1.4cm}\centering
\epsfig{width=.45\textwidth,file=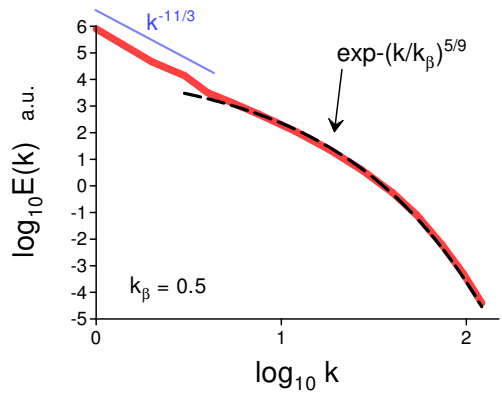} \vspace{-3.7cm}
\caption{The depth- and time-averaged barotropic kinetic energy spectrum for pure hydrodynamic rapidly rotating Rayleigh-B\'{e}nard convection (${\tilde Q} =0$).}
\end{figure}
   Figure 15 shows kinetic energy spectrum of the barotropic convective eddies' background (the small-scale part of the barotropic spectrum) at the end  of the computations ($t=100$ in the terms of the Ref. \cite{rub}). The spectral data were taken from Fig. 3 of the Ref. \cite{rub}. The dashed curve indicates the stretched exponential spectral law Eq. (19) (i.e. the distributed chaos, see also below).\\

  In recent paper Ref. \cite{ede} a DNS of the core convection zone of the massive stars were performed using an anelastic approximation in a 3D sphere. Since the equations in this approximation are not different, in respect of applicability of the Sections II and III, from the Eqs. (20-22) we can apply considerations of these Sections to the results obtained in the DNS. \\
  
  Figure 16 shows kinetic energy spectrum obtained for a non-rotational DNS in the convection zone at the Rayleigh number $Ra = 10^{12}$ and the Pandtl number $Pr = 100$. The spectral data were taken from Fig. 9 of Ref. \cite{ede} (H6LD run). In the Ref. \cite{ede} the spectrum is given against spherical harmonic degree $l$. On a sphere of radius $R$ the wavenumber $k$ is related to $l$ as 
$$
k = \sqrt{l(l+1)}/R  \eqno{(24)}
$$
  
   In the Fig. 16 the spectrum is given against $k$ and recalculated, according to the relationship Eq. (24), for $R =1$. The dashed curve in the Fig. 16 indicates the stretched exponential law Eq. (17). \\

  Figure 17 shows kinetic energy spectrum obtained for a rotational DNS at the angular velocity $\Omega = 10^{-5}$ rad/s in the convection zone at the Rayleigh number $Ra = 8 \times 10^{11}$, and the Pandtl number $Pr = 60$. In the Ref. \cite{ede} a rotating frame of reference was used with rotation axis in the direction of the pole. The spectral data were taken from Fig. 9 of ref. \cite{ede} (H6R10 run). The dashed curve in the Fig. 17 indicates the stretched exponential law Eq. (19).\\
  
 \subsection{Rayleigh-Taylor convection}
 
   In the gravity field the Rayleigh-Taylor instability occurs at the the interface between regions of different density (temperature). At this instability a light fluid pushes on a heavy fluid progressively resulting in a chaotic and turbulent motion in a mixing zone which grows in time (see for a review Ref. \cite{abar}). \\
   
   The Eqs. (20-22) were used in the Ref. \cite{bmm} for a DNS of such process in a periodic spatial volume $L_x=L_y=L_z/4$. The initial conditions were: ${\bf u }({\bf x}, 0) =0$, $T({\bf x}, 0) =-(\Delta T/2) sgn (z)$, $\Delta T$ is the temperature jump at the interface (plane) $z=0$. Corresponding Atwood number is defined as $A = \sigma \Delta T/2$. In the DNS the Atwood number $A = 0.25/g$, the Prandtl number $Pr = 1$, $\Delta T =1$. To initiate the Rayleigh-Taylor instability a weak white noise was added to the initial density field in a vicinity of the plane $z=0$.\\
   
   It is observed that the imposed rotation along the vertical axis results in qualitative and quantitative changes in the Rayleigh-Taylor convection: the thermal plumes became elongated in the vertical direction and more coherent, whereas there vertical velocity decreased. \\
   
    Figure 18 shows the power spectrum of the horizontal component of the velocity - $E_x$, against the wavenumber $k = \sqrt{k_x^2+k_y^2+k_z^2}$ at $\Omega =0$. The data were taken from Fig. 9a of the Ref. \cite{bmm}. The dashed curve in the Fig. 18 indicates the stretched exponential law Eq. (19).\\

\begin{figure} \vspace{-1.5cm}\centering
\epsfig{width=.45\textwidth,file=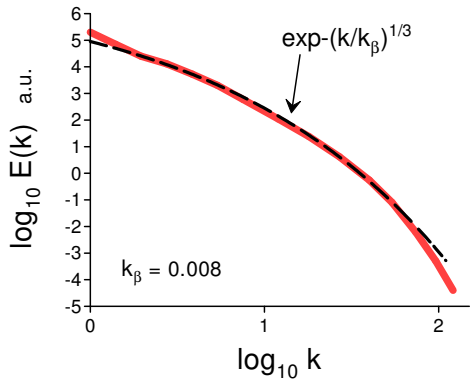} \vspace{-3.7cm}
\caption{As in the Fig. 21 but for rapidly rotating magnetohydrodynamic Rayleigh-B\'{e}nard convection with an imposed mean magnetic field at ${\tilde Q} =0.1$.}
\end{figure}
\begin{figure} \vspace{-0.5cm}\centering
\epsfig{width=.45\textwidth,file=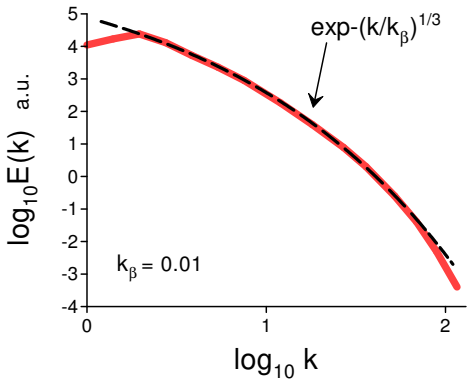} \vspace{-3.7cm}
\caption{As in the Fig. 21 but for the rapidly rotating magnetohydrodynamic Rayleigh-B\'{e}nard convection with an imposed mean magnetic field at ${\tilde Q} =1$.}
\end{figure}

     Figure 19 shows the energy spectrum of the horizontal component of the velocity - $E_x$, at $\Omega \tau =12.5$ (where the characteristic time $\tau = (L_x/Ag)^{1/2}$). The data were taken from Fig. 9a of the Ref. \cite{bmm}. The dashed curve in the Fig. 19 indicates the stretched exponential law Eq. (19). \\
     
     Figure 20 shows analogous power spectrum for the vertical component of the velocity at $\Omega \tau =12.5$. The data were also taken from Fig. 9a of the Ref. \cite{bmm}. The dashed curve in the Fig. 20 indicates the stretched exponential law Eq. (19).
     
 \subsection{The rotating Rayleigh-B\'{e}nard convection in magnetohydrodynamics}

   In a recent paper Ref. \cite{maf} the rapidly rotating Rayleigh-B\'{e}nard convection has been numerically studied also in a magnetohydrodynamic version for an electrically conducting fluid. The hydrodynamic approach was similar to that used in the above-quoted Ref. \cite{rub}. In the pure hydrodynamic version the formation of the large-scale barotropic vortices was also observed, although the formation of the large-scale vortices is suppressed by a sufficiently strong magnetic field.\\
   
    Figure 21 shows the depth- and time-averaged power spectrum of the barotropic velocity observed in the pure hydrodynamic version (cf. Fig. 15, where the only small-scale convective eddies' background is shown). The data were taken from Fig. 3 of the Ref. \cite{maf}. The data corresponds to the asymptotically-scaled Rayleigh number $\tilde{Ra}= Ra~E^{4/3} = 160$, where the $E$ is the Ekman number. The dashed curve indicates the stretched exponential spectral law Eq. (19) (as in the Fig. 15). \\
    
    The present DNS were performed in a horizontally-periodic layer of depth $L$, ${\boldsymbol \Omega} = \Omega {\bf e}_z$. The boundary conditions were stress-free, impenetrable and fixed-temperature. For the magnetohydrodynamic (MHD) version the boundary conditions were perfectly electrically conducting. In the MHD version a horizontal mean magnetic field 
 $$
{\bf B_0} = \frac{\sqrt{2}}{2}\{ [\cos(\pi Z) - \cos(3\pi Z)]{\bf e}_x - \cos(\pi Z){\bf e}_y \} \eqno{(25)},
$$
(where $Z = E^{1/3} z$) was imposed on an electrically conducted fluid. The magnetic field in the system was characterized by the asymptotically-scaled Chandrasekhar number ${\tilde Q} = Q E^{4/3}$, where $Q = \mathcal{B}^2 L^2/\mu_0 \rho\nu\eta$, $\mathcal{B}$ is the magnitude of the mean magnetic field, $\mu_0$ is the free-space magnetic permeability and $\eta$ is the magnetic diffusivity.  \\

  In the MHD version the Eq. (23) should be replaced by equation taking into account the Lorentz force
$$
\frac{d\langle h \rangle}{dt}  = 2\langle {\boldsymbol \omega}\cdot (-2 {\boldsymbol \Omega} \times {\bf u} + \sigma g T {\bf e}_z -  [{\bf b} \times (\nabla \times {\bf b})])\rangle \eqno{(26)} 
$$ 
where the normalized magnetic field ${\bf b} = {\bf B}/\sqrt{\mu_0\rho}$ (in these Alfv\'enic units the magnetic field has the same dimension as velocity).  All the considerations of the Sections II and III can be applied to this case as well.    \\

  Figures 22 and 23 show the depth- and time-averaged barotropic kinetic energy spectrum for the rapidly rotating magnetohydrodynamic Rayleigh-B\'{e}nard convection with the imposed mean magnetic field at ${\tilde Q} = 0.1$ and ${\tilde Q} = 1$ respectively. The data were taken from Fig. 3 of the Ref. \cite{maf}. The dashed curve indicates the stretched exponential spectral law Eq. (18) in these figures.

\subsection{The Earth's crustal magnetic field} 

The directly measured geomagnetic field is a superposition of the main field, presumably generated by the geodynamo in the outer (liquid) core and the magnetic field of magnetized rocks in the crust. The main field dominates the comparatively long wavelengths, whereas the crustal geomagnetic field originating from magnetized crustal rocks dominates the geomagnetic power spectrum at the wavelengths between 0.1 and 100 km \cite{lg}.  \\

\begin{figure} \vspace{-1.9cm}\centering
\epsfig{width=.45\textwidth,file=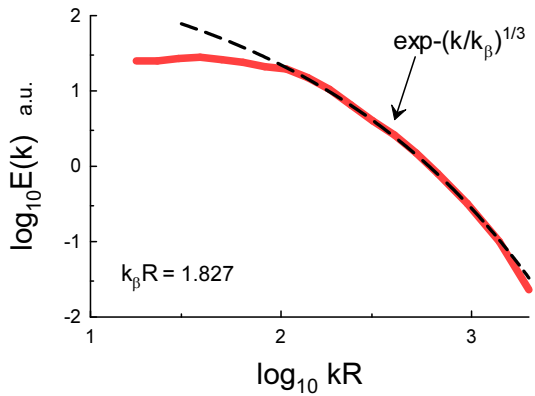} \vspace{-4.2cm}
\caption{Crustal magnetic power spectrum computed \\ 
using the World Magnetic Anomaly Map.} 
\end{figure}

  The crustal magnetic filed is composed from induced and remanent magnetisation. The remanent (paleo-) magnetism, is
a permanent magnetism in rocks, originating at the time of the rocks formation. The remanent magnetisation dominates the comparatively short wavelength and it is, naturally, strong near the Earth surface.\\

  On the other hand, the short-wavelength part of the crustal geomagnetic field is closely correlating with the near-surface geologic variations (it is especially strong in the vicinity of magnetic and ferrous geological materials). Therefore, it is not clear whether the remanent (paleo-) magnetism in the Earth crust will be globally dominating factor at the short wavelengths. If we assume a random distribution of the geologic variations on a global map, then this hypothesis can be readily verified. \\
  
\begin{figure} \vspace{-0.8cm}\centering \hspace{-1cm}
\epsfig{width=.48\textwidth,file=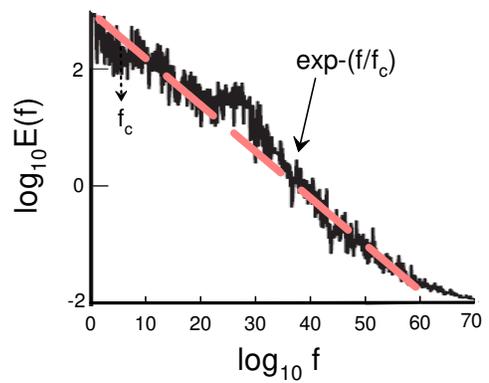} \vspace{-5.03cm}
\caption{Frequency spectrum of kinetic energy for the \\ 
time-asymptotic states at the Rayleigh number $Ra =1.65\times10^5$.} 
\end{figure}

  Let us recall that in the Alfv\'enic units ${\bf b} = {\bf B}/\sqrt{\mu_0\rho}$ the magnetic field has the same dimension as velocity. Therefore, in the Alfv\'enic units the same dimensional considerations, which were used to obtain the velocity power spectrum Eq. (18), can be used to obtain the power spectrum of the magnetic field in the form of the Eq. (18). Figure 24 shows geomagnetic power spectrum computed using subgrids ($20^{o}\times 20^o$) of the National Geophysics Data Center's (NGDC's) World Magnetic Anomaly Map (WMAM \cite{maus1}). The mean field was subtracted, and the data were detrended and azimuthally averaged (the spectral data were taken from Fig. 4 of the Ref. \cite{maus2}).

  The dashed curve in the Fig. 24 indicates correspondence to the spectral law Eq. (18). The $k_{\beta}R \simeq 1.827$ ($R = 6,371$ km is the mean Earth's radius).  Hence $1/k_{\beta} \simeq 3,487$ km. Thus the characteristic spatial scale $1/k_{\beta}$ is approximately equal to the radius of the outer (liquid) core of the Earth $R_0 \simeq 3,483 \pm 5$ km \cite{ahr}, as one can expect for the geomagnetic dynamo.  

\subsection{Temporal variability of the geomagnetic dynamo}

  In previous section we considered spatial variability of the geomagnetic field. Now let us turn to temporal variability of the paleo-geomagnetic dynamo. \\
  
    In paper Ref. \cite{feu} a route to chaos for an incompressible fluid in a rotating spherical shell driven by a central gravity field and heated from the inner boundary
(the situation characteristic for the geomagnetic dynamo situation) was studied using direct numerical simulations. It was found that for the Rayleigh number $Ra =1.65\times 10^5$ the fluid motion can be already considered as a developed deterministic chaos. An exponential spectrum of the computed kinetic energy was presented in the Ref. \cite{feu} as an indication that the motion at this Rayleigh number was at the state of deterministic chaos. Figure 25 shows this spectrum (the spectral data were taken from Fig. 11 of the Ref. \cite{feu}). The dashed straight line indicates the exponential frequency spectrum
$$
E(f) \propto \exp-(f/f_c)   \eqno{(27)}
$$
  cf. the spatial (wavenumber) exponential spectrum Eq. (3). The characteristic frequency $f_c$ becomes fluctuating with increasing Rayleigh number and the chaos is transforming from the deterministic into the distributed one. The above consideration of the spatial distributed chaos can be applied to the temporal distributed chaos as well.\\

  The temporal variability of the geomagnetic dynamo is usually studied using the temporal variability of the geomagnetic dipole moment. The geomagnetic dipole moment (normalized by the volume of motion - $V$) is
$$
{\boldsymbol \mu} =   \frac{1}{2V} \int [{\bf r} \times {\bf j}]~ dV = \frac{1}{2V} \int [{\bf r} \times (\nabla \times {\bf b})]~ dV  \eqno{(28)}
$$

\begin{figure} \vspace{+0.26cm}\centering \hspace{-1cm}
\epsfig{width=.515\textwidth,file=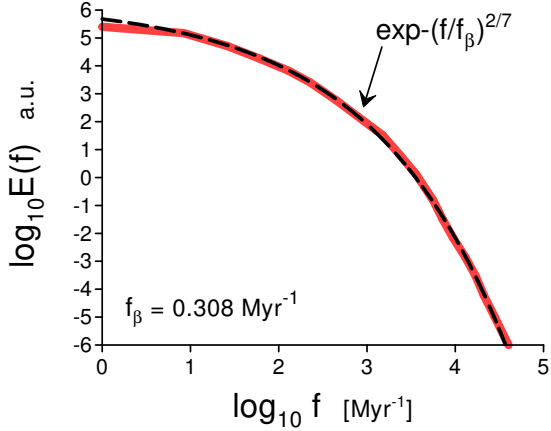} \vspace{-4.9cm}
\caption{Power spectrum of the axial dipole magnetic \\ 
field (numerical simulation, $Rm = 130$ and $Ra \simeq 1.08\times 10^7$).} 
\end{figure}
\begin{figure} \vspace{+0.2cm}\centering \hspace{-1cm}
\epsfig{width=.5\textwidth,file=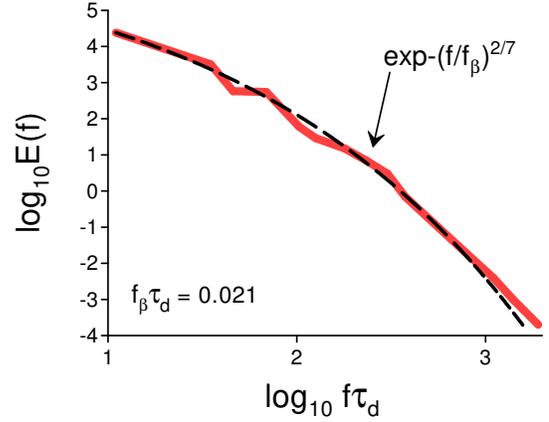} \vspace{-4.5cm}
\caption{Power spectrum of the axial dipole magnetic \\ 
field (numerical simulation, $Rm = 1985$ and $Ra = 5\times 10^9$).} 
\end{figure}
\begin{figure} \vspace{-0.5cm}\centering \hspace{-0.6cm}
\epsfig{width=.515\textwidth,file=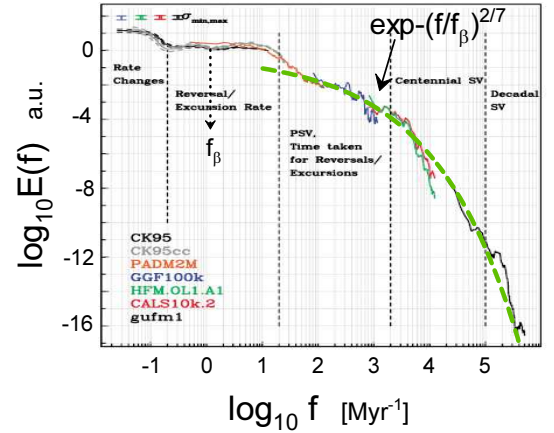} \vspace{-5.34cm}
\caption{Extended composite power spectrum of geomagnetic dipole moment variations for the time interval 0-160 Myr.}
\end{figure}

 This normalized geomagnetic dipole moment in the Alfv\'enic units ${\bf b} = {\bf B}/\sqrt{\mu_0\rho}$ has the same dimension as velocity, and the same type of dimensional considerations can be used in this case as well. Now, however,  the spatial (wavenumber) estimate Eq. (13) should be replaced by the temporal (frequency) dimensional estimate 
 $$
 b_c \propto  |I_n|^{1/(4n-3)}~ f_c^{\alpha_n}    \eqno{(29)}
$$
  Hence, in this case
 $$
 \alpha_n = \frac{2n-3}{4n-3}   \eqno{(30)}
 $$
 and from Eq. (15) we obtain 
 $$
 \beta_n = \frac{ 2(2n-3)}{(8n-9)}  \eqno{(31)}
 $$
 
 It follows from the Eq. (31) that for $n \gg 1$ the $\beta_n \simeq 1/2$, i.e.
 $$
 E(f) \propto \exp(-f/f_{\beta})^{1/2}   \eqno{(32)}
 $$
 and for $n=2$ the $\beta_n =2/7$, i.e. 
 $$
 E(f) \propto \exp(-f/f_{\beta})^{2/7}   \eqno{(33)}
$$

   In paper Ref. \cite{dc} a standard system of equations for the geomagnetic dynamo driven by the thermal convection (in the Boussinesq approximation) in a shell between two concentric spheres rotating with a global angular velocity $\Omega$ and appropriate boundary conditions was used for direct numerical simulations.\\

   Figure 26 shows a power spectrum of the axial dipole magnetic field computed at these simulations for the magnetic Reynolds number $Rm = 130$ and the Rayleigh number $Ra \simeq 1.08\times 10^7$ (the spectral data for the Fig. 26 were taken from Fig. 3 of the Ref \cite{dc}). The dashed curve is drawn to indicate the spectrum Eq. (33). \\
   
   Analogous DNS were reported in Ref. \cite{ocd}. Figure 27 shows a power spectrum of the axial dipole magnetic field computed at these simulations for the magnetic Reynolds number $Rm = 1985$ and the Rayleigh number $Ra =5\times 10^9$ (the spectral data for the Fig. 27 were taken from Fig. 2 of the Ref \cite{ocd}). In this figure the frequency was scaled with the dipole free decay time $\tau_d = r_0^2/\pi^2 \eta$, where $r_0$ is the outer radius of the shell and $\eta$ is the magnetic diffusivity. The dashed curve is drawn to indicate the spectrum Eq. (33). \\
   
    Figure 28 shows power spectrum of the geomagnetic dipole moment composed from the CK95 reversal record (0-160 Myr), the CK95cc reversal record cryptochrons (0-83 Myr), the PADM2M axial dipole moment reconstruction (0-2 Myr), the HFM.OL1.A1 and the CALS10k.2 Holocene geomagnetic field models, the gufm1 historical geomagnetic field model, and the GGF100k geomagnetic field model (0-100kyr). The thick dashed curve is drawn to indicate correspondence to the spectrum Eq. (33), and the dotted arrow indicates position of the $f_{\beta}$ (the spectral data were taken from Fig. 8 of the Ref. \cite{pck}).

\end{document}